\documentclass[a4paper]{article}
\usepackage[margin=0.70in]{geometry}
\usepackage{amsmath,amssymb,empheq}
\usepackage{amsfonts} 
\usepackage{graphicx}
\usepackage{parskip} 
\usepackage{subcaption} 
\usepackage{multirow}	
\usepackage{float} 
\usepackage{siunitx}
\usepackage{comment}
\usepackage{todonotes}
\usepackage{mathtools}

\usepackage{pgfplotstable}
\usepackage{pgfplots}

\usepackage{algorithm}
\usepackage{algpseudocode}

\usepackage{xcolor}

\usepackage{authblk}

\usepackage{adjustbox}

\usepackage{url}

\usepackage{enumitem}

\newcounter{assctr}
\newcounter{remctr}

\newenvironment{assumption}{
   \refstepcounter{assctr}
   \textit{Assumption \theassctr}
   \newline
   \hangindent=0.7cm%
   }{\par}  

\newenvironment{remark}{
   \refstepcounter{remctr}
   \textbf{Remark \theremctr.}
   }{\par}  

\newcounter{exctr}

\newenvironment{example}[1]{
    \refstepcounter{exctr}
    \textit{Example \theexctr: #1}
    \newline
    \hangindent=0.7cm%
    }{\par}

\title{Cell Escape Probabilities for Markov Processes on a Grid}

\author[1]{T. Ingelaere\thanks{Corresponding Author: toon.ingelaere@kuleuven.be}}
\author[1]{V. Maes}
\author[1]{G. Samaey}
\affil[1]{NUMA, Department of Computer Science, KU Leuven, Belgium}

\date{}

\begin{document}
\pagenumbering{arabic}
\setcounter{page}{1}
\maketitle

\section*{Abstract}
Kinetic equations describe physical processes in a high-dimensional phase space and are often simulated using Markov process-based Monte Carlo routines. The quantities of interest are typically defined on the lower-dimensional position space and estimated on a grid (histogram). In several applications, such as the construction of diffusion Monte Carlo-like techniques and variance prediction for particle tracing Monte Carlo methods, the cell escape probabilities, i.e., the probabilities with which particles escape a grid cell during one step of the Markov process, are of interest. In this paper, we derive formulas to calculate the cell escape probabilities for common mesh elements in one, two, and three dimensions. Deterministic calculation of cell escape probabilities in higher dimensions becomes expensive and prone to quadrature errors due to the involved high-dimensional integrals. We therefore also introduce a stochastic Monte Carlo algorithm to calculate the escape probabilities, which is more robust at the cost of a statistical error. The code used to perform the numerical experiments and accompanying GeoGebra tutorials are openly available at \texttt{https://gitlab.kuleuven.be/numa/public/escape-probabilities-markov-processes}.

Keywords: escape probability, kinetic equation, Monte Carlo, mesh elements, particle tracing, diffusion Monte Carlo, variance prediction

\section{Introduction}
\label{sec:introduction}
Large particle systems are typically described by a high-dimensional kinetic equation
\begin{equation}
    \partial_t f(\boldsymbol{x},\boldsymbol{s}, t) = F(f(\boldsymbol{x},\boldsymbol{s},t),\boldsymbol{x},\boldsymbol{s},t).
    \label{eq:general_kinetic_equation}
\end{equation}
The particle distribution is denoted by $f(\boldsymbol x, \boldsymbol s, t)$, where $\boldsymbol x \in \Omega \subseteq \mathbb{R}^n$ represents the particle position (typically $n \in \{1,2,3\}$), $\boldsymbol s \in \mathbb{R}^m$ represents other state variables (e.g., the particle velocity $\boldsymbol v$) in a high-dimensional phase space, and $t \geq 0$ represents time. A kinetic equation can be interpreted as describing the macroscopic population dynamics corresponding to some microscopic particle dynamics, which often turn out to be a Markov process: the next particle position $\boldsymbol x_{k+1}$ and state $\boldsymbol s_{k+1}$ depend on the current position $\boldsymbol x_k$ and state $\boldsymbol s_k$, but not on prior ones:
\begin{alignat}{2}
    \boldsymbol x_{k+1} & = \boldsymbol x_k + \Delta\boldsymbol x_k, \qquad & \Delta\boldsymbol x_k & \sim p(\Delta\boldsymbol x | \boldsymbol x_k,\boldsymbol s_k,t_k), \label{eq:markov_process} \\
    \boldsymbol s_{k+1} & = \boldsymbol s_k + \Delta\boldsymbol s_k, \qquad & \Delta\boldsymbol s_k & \sim p(\Delta\boldsymbol s | \boldsymbol x_k,\boldsymbol s_k,t_k).
\end{alignat}
The Wiener process and velocity-jump processes are two examples of such Markov processes~\cite{lapeyre_introduction_2003}.

\begin{example}{The $n$-dimensional Wiener process}
    The Wiener process sampled at equidistant times $t_k$ is characterized by independent, zero-mean, Gaussian increments
    \begin{equation}
        \Delta\boldsymbol x_k \sim N(0, \Delta tI_n),
    \end{equation}
    where $\Delta t$ denotes the time interval between samples. The step $\Delta\boldsymbol x_k$ is independent of $\boldsymbol x_k$ and $t_k$ and there are no additional state variables $\boldsymbol s$. The population dynamics corresponding to this Wiener process are described by the kinetic equation
    \begin{equation}
        \partial_t f(\boldsymbol x, t) = \frac{1}{2}\nabla^2f(\boldsymbol x, t).
    \end{equation}\label{ex:wiener}
\end{example}
\begin{example}{A simple velocity-jump process}
    Consider a simple example of an $n$-dimensional velocity-jump process where a particle updates its velocity $\boldsymbol v_k$ (a state variable) at random jump times $t_k$. The jump times follow a Poisson process. The travel time between jumps $\Delta t_k$ is therefore exponentially distributed:
    \begin{equation}
        \Delta t_k \sim \lambda \exp(-\lambda \Delta t), \quad \Delta t\geq 0.
    \end{equation}
    The velocities $\boldsymbol v_k$ are sampled independently from $N(0,I_n)$. As a result, we have
    \begin{equation}
        \Delta \boldsymbol v_k \sim p(\Delta\boldsymbol v|\boldsymbol v_k) = N(-\boldsymbol v_k, I_n).
    \end{equation}
    The steps between jumps are then sampled from
    \begin{equation}
        \Delta \boldsymbol x_k = \boldsymbol v_k \Delta t_k \sim \int_0^\infty \frac{1}{\Delta t^n}\lambda \exp(-\lambda \Delta t)\frac{\exp\left(-\frac{\Delta \boldsymbol x^T \Delta \boldsymbol x}{2\Delta t^2}\right)}{\sqrt{(2\pi)^n}}\;\mathrm{d} (\Delta t).
    \end{equation}
    The kinetic equation describing the population dynamics corresponding to this velocity jump process is
    \begin{equation}
        \partial_t f(\boldsymbol x, \boldsymbol v, t) = - \boldsymbol v \cdot \nabla_{\boldsymbol x} f(\boldsymbol x, \boldsymbol v, t) - \lambda f(\boldsymbol x, \boldsymbol v, t) + \lambda \frac{\exp\left(-\frac{\boldsymbol v^T \boldsymbol v}{2}\right)}{\sqrt{(2\pi)^n}} \int\limits_{\mathbb R^n} f(\boldsymbol x, \boldsymbol v',t)\;\mathrm{d}\boldsymbol v'.
    \end{equation}
\end{example}

Quantities of interest are typically moments of the particle distribution $f(\boldsymbol x,\boldsymbol s,t)$ of the form
\begin{equation}
    Q(\boldsymbol{x},t) = \int q(\boldsymbol{s})f(\boldsymbol{x},\boldsymbol{s},t)\ \mathrm{d}\boldsymbol{s},
    \label{eq:QoI_Generic}
\end{equation}
where $q(\boldsymbol s)$ is a function of the state variables. We focus on the setup where simulation of the Markov process and estimation of a quantity of interest happen on a discretization of the domain $\Omega$ into a grid of cells $\{\Omega_j\}_{j=1}^J$.

For several applications, it is desirable to have a uniform framework for calculating the cell escape probability ${P(\boldsymbol x_{k+1} \notin \Omega_j | \boldsymbol x_k \in \Omega_j)}$ for some grid cell $\Omega_j \subset \Omega$, specifically in the context where both the step $\Delta \boldsymbol x_k$, but also the initial position of the particle $\boldsymbol x_k$ are random. A first application is variance prediction for quantity-of-interest estimation using particle tracing Monte Carlo methods~\cite{Vince}. In that field, the cell escape probability is an important parameter for deriving a local Markov process-based variance predictor for a cell. Second, the framework for calculating the escape probability from a grid cell can be extended to the derivation of transition probabilities between two grid cells $P(\boldsymbol x_{k+1} \in \Omega_i | \boldsymbol x_k \in \Omega _j)$, which can be used for hidden Markov process-based variance calculations~\cite{Vince}.  A third application is to use the escape probabilities or the related transition probabilities for the construction of diffusion Monte Carlo-like algorithms~\cite{urbatsch_monte_1999, evans_1-D_2000, densmore_interface_2006, densmore_hybrid_2007, densmore_hybrid_2012} where these probabilities allow to quantify the leakage of particles from their current cell to one of the other (neighbouring) cells in the grid.

In this paper, we derive deterministic formulas for cell escape probabilities from several common one-, two-, and three-dimensional mesh elements. In the one-dimensional case, we consider line elements. In two dimensions, we consider triangular and parallelogrammatic elements. In three dimensions, we consider tetrahedral and parallelepipedal elements. Our formulas are derived under two core assumptions: given $\boldsymbol x_k\in\Omega_j$, we presuppose both uniformity of $\boldsymbol x_k$ and independence between $\boldsymbol x_k$ and $\Delta\boldsymbol x_k$. We also introduce a Monte Carlo algorithm for calculating the escape probabilities, which can be used in the case where these two core assumptions are not satisfied or where deterministic computation of the involved integrals becomes prohibitive. For an $n$-dimensional Wiener process, we illustrate numerically that the deterministic formulas are faster and more accurate than the Monte Carlo alternative.

\textbf{Overview of the paper.} The paper is structured as follows. Section~\ref{sec:general_approach} provides the general approach to obtain expressions for the escape probabilities from arbitrary geometries. These expressions involve two independent factors. The first factor is the step probability, which depends on the governing particle dynamics. The second factor is the so-called conditional escape probability, which depends on the considered geometry. Expressions for the conditional escape probability for common mesh elements are derived in Section~\ref{sec:cond_escp_prob}. Section~\ref{sec: mca} introduces an alternative Monte Carlo algorithm for calculating cell escape probabilities. In Section~\ref{sec:numerical_results}, we perform several numerical experiments to corroborate the obtained escape probability expressions and quantify the computational cost. A short conclusion and some pointers for future work are given in Section~\ref{sec: conclusion}. The Appendix contains an extension of the framework from escape probabilities to the computation of transition probabilities. The code and GeoGebra tutorials accompanying this paper are openly available at~\cite{code}.

\section{General Computation of the Escape Probability}
\label{sec:general_approach}
Consider a Markov process~\eqref{eq:markov_process} in which a particle additively updates its position $\boldsymbol x_k\in \mathbb R^n$ with a step $\Delta\boldsymbol x_k \in \mathbb R^n$. Both  $\boldsymbol x_k \sim p(\boldsymbol x| t_k)$ and $\Delta\boldsymbol x_k \sim p(\Delta\boldsymbol x| \boldsymbol x_k, \boldsymbol s_k, t_k)$ are random variables. For the remainder of this section, we will drop the dependence on $t_k$ and $\boldsymbol s_k$ in our notation. The goal is to determine the \textit{escape probability} ${P(\boldsymbol x_{k+1} \notin \Omega_j | \boldsymbol x_k \in \Omega_j)}$. This is the probability that a particle exits a cell $\Omega_j \subset \Omega$ by additively updating its position $\boldsymbol x_k$ with step $\Delta\boldsymbol x_k$. To do so, we make the following two core assumptions.

\begin{assumption}
    The particle position $\boldsymbol x_k$ is uniformly distributed within the cell $\Omega_j$ from which we want to calculate the escape probability: \label{ass: uniform_x}
    \begin{equation}
        p(\boldsymbol x) = c, \quad \forall \boldsymbol x\in\Omega_j.
    \end{equation}
\end{assumption}
\begin{assumption}
    Within $\Omega_j$, the step $\Delta \boldsymbol x_k$ is independent of the current position $\boldsymbol x_k$. For any $\boldsymbol x \in \Omega_j$, the step $\Delta\boldsymbol x_k$ is sampled from the same distribution $p(\Delta\boldsymbol x)$:\label{ass: ind_step}
    \begin{equation}
        p(\Delta\boldsymbol x|\boldsymbol x) = p(\Delta\boldsymbol x), \quad \forall \boldsymbol x\in\Omega_j.
    \end{equation}
\end{assumption}
These assumptions correspond to the kinetic equation~\eqref{eq:general_kinetic_equation} being homogeneous in $\Omega_j$, i.e., having constant spatial parameters and a constant solution in $\Omega_j$. These are common assumptions for grid cells in, for example, a finite volume discretization~\cite{leveque_finite_2002}.

\begin{remark}
    If both Assumption \ref{ass: uniform_x} and \ref{ass: ind_step} hold, then the joint probability distribution of $\boldsymbol x_k$ and $\Delta\boldsymbol x_k$ for $\boldsymbol x\in\Omega_j$ can be written as
    \begin{equation}
        p(\boldsymbol x, \Delta\boldsymbol x) = p(\boldsymbol x)p(\Delta\boldsymbol x|\boldsymbol x) = cp(\Delta\boldsymbol x), \quad \forall \boldsymbol x\in\Omega_j.
        \label{eq:ass1and2}
    \end{equation}
    Note that no assumptions are made on the behavior of $p(\boldsymbol x, \Delta\boldsymbol x)$ outside of $\Omega_j$, because to compute the escape probability, we presuppose that $\boldsymbol x_k \in\Omega_j$.
\end{remark}
\begin{remark}
    When particles can reach and interact with a boundary of the domain $\Omega$, Assumption \ref{ass: ind_step} is violated. The possible interactions of a particle with a domain boundary are reflected in the step probability density $p(\Delta\boldsymbol x|\boldsymbol x_k)$, which is then no longer independent of $\boldsymbol x_k$, because the distance from a particle to a domain boundary depends on the specific value of its position.
\end{remark}

\subsection{Integral Formulation}\label{sec: integral}
By the law of total probability, we can write the escape probability as an integral over $\Delta\boldsymbol x$:
\begin{equation}
    \begin{split}
        P(\boldsymbol x_{k+1} \notin \Omega_j | \boldsymbol x_k \in \Omega_j) &= \int\limits_{\mathbb R^n} P(\boldsymbol x_{k+1} \notin \Omega_j | \boldsymbol x_k \in \Omega_j, \Delta\boldsymbol x)p(\Delta\boldsymbol x | \boldsymbol x_k\in\Omega_j)\;\mathrm d(\Delta\boldsymbol x)\\
        &= \int\limits_{\mathbb R^n} P(\boldsymbol x_{k+1} \notin \Omega_j | \boldsymbol x_k \in \Omega_j, \Delta\boldsymbol x)p(\Delta\boldsymbol x)\;\mathrm d(\Delta\boldsymbol x),
    \end{split}
    \label{eq:integral_formulation}
\end{equation}

where Assumption \ref{ass: ind_step} is used for the second equality. The integrand consists of two factors. We call the first factor, $P(\boldsymbol x_{k+1} \notin \Omega_j | \boldsymbol x_k \in \Omega_j,\Delta\boldsymbol x_k)$, the \textit{conditional escape probability}, given $\Delta\boldsymbol x_k$. The conditional escape probability is the probability that a particle exits the cell $\Omega_j$ for a given step $\Delta\boldsymbol x_k$. Given Assumptions \ref{ass: uniform_x} and \ref{ass: ind_step}, the conditional escape probability depends solely on the shape and size of cell $\Omega_j$. We call the second factor, $p(\Delta\boldsymbol x)$, the step probability density. The step probability density solely depends on the kinetic equation~\eqref{eq:general_kinetic_equation} that governs the particle dynamics. In this paper, we focus on deriving expressions for the conditional escape probability, specifically in the case where $\Omega_j$ represents a common mesh element. If we can compute the conditional escape probability for a general $\Delta\boldsymbol x$, then we can multiply it with the step probability and integrate over all possible values of $\Delta\boldsymbol x$ to obtain the sought-after escape probability.

\subsection{General Computation of the Conditional Escape Probability}
\begin{figure}
    \centering
    \includegraphics[width=0.4\linewidth]{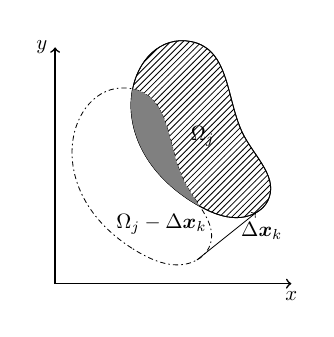}
    \caption{Cell $\Omega_j$ and its translation $\Omega_j-\Delta\boldsymbol x_k$. If Assumptions \ref{ass: uniform_x} and \ref{ass: ind_step} hold, then the conditional escape probability for a given $\Delta\boldsymbol x_k$ corresponds to the proportion of $\Omega_j$ which is hatched.}
    \label{fig: translation}
\end{figure}

To compute the conditional escape probability $P(\boldsymbol x_{k+1} \notin \Omega_j | \boldsymbol x_k\in \Omega_j, \Delta\boldsymbol x_k)$, we first perform the following manipulations:
\begin{equation}
    \begin{split}
        P(\boldsymbol x_{k+1} \notin \Omega_j | \boldsymbol x_k\in \Omega_j, \Delta\boldsymbol x_k) &= \frac{P(\boldsymbol x_k\in \Omega_j\ \mathrm{and}\ \boldsymbol x_{k+1}\notin\Omega_j|\Delta\boldsymbol x_k)}{P(\boldsymbol x_k\in\Omega_j|\Delta\boldsymbol x_k)}\\
        &= \frac{P(\boldsymbol x_k\in \Omega_j\ \mathrm{and}\ \boldsymbol x_k+\Delta\boldsymbol x_k\notin\Omega_j|\Delta\boldsymbol x_k)}{P(\boldsymbol x_k\in\Omega_j|\Delta\boldsymbol x_k)}\\
        &= \frac{P(\boldsymbol x_k\in \Omega_j\ \mathrm{and}\ \boldsymbol x_k\notin(\Omega_j-\Delta\boldsymbol x_k)|\Delta\boldsymbol x_k)}{P(\boldsymbol x_k\in\Omega_j|\Delta\boldsymbol x_k)}\\
        &= \frac{P(\boldsymbol x_k\in \Omega_j\setminus(\Omega_j-\Delta\boldsymbol x_k)|\Delta\boldsymbol x_k)}{P(\boldsymbol x_k\in\Omega_j|\Delta\boldsymbol x_k)},\\
    \end{split}
    \label{eq:conditional_escape_prob_step1}
\end{equation}

where $\Omega_j - \Delta\boldsymbol x_k = \{\boldsymbol x -\Delta\boldsymbol x_k\in \mathbb R^n | \boldsymbol x \in \Omega_j\}$ is the image of $\Omega_j$ under translation by $-\Delta\boldsymbol x_k$, as visualized in Figure~\ref{fig: translation}.

In the derivation of equation~\eqref{eq:conditional_escape_prob_step1}, we first used the definition of conditional probability. Subsequently, we substituted $\boldsymbol x_{k+1} = \boldsymbol x_k+\Delta \boldsymbol x_k$. Finally, the definition of the difference of two sets is used. In the end, the conditional escape probability equals the proportion of particles in $\Omega_j$ that are not in $\Omega_j - \Delta\boldsymbol x_k$, i.e., particles in the hatched area in Figure~\ref{fig: translation}. Indeed, these particles leave $\Omega_j$ when translated by $\Delta\boldsymbol x_k$.

We introduce the symbol $V(A)$ to denote either the length, area or volume of a set $A$, depending on its dimensionality. Under Assumptions \ref{ass: uniform_x} and \ref{ass: ind_step}, we can use~\eqref{eq:ass1and2} to calculate the numerator and denominator in~\eqref{eq:conditional_escape_prob_step1} as integrals over the joint probability density function:
\begin{equation}
    \begin{split}
        P(\boldsymbol x_k \in \Omega_j \setminus (\Omega_j-\Delta\boldsymbol x_k)|\Delta\boldsymbol x_k) & = \int\limits_{\mathclap{\Omega_j\setminus(\Omega_j-\Delta\boldsymbol x_k)}}\frac{p(\boldsymbol x,\Delta\boldsymbol x_k)}{p(\Delta\boldsymbol x_k)}\;\mathrm d\boldsymbol x \\
        & = cV(\Omega_j \setminus (\Omega_j-\Delta\boldsymbol x_k)),
    \end{split}
\end{equation}

\begin{equation}
    \begin{split}
        P(\boldsymbol x_k \in \Omega_j|\Delta\boldsymbol x_k) & = \int\limits_{\Omega_j}\frac{p(\boldsymbol x,\Delta\boldsymbol x_k)}{p(\Delta\boldsymbol x_k)}\;\mathrm d\boldsymbol x \\
        & = cV(\Omega_j).
    \end{split}
\end{equation}
The conditional escape probability~\eqref{eq:conditional_escape_prob_step1} now reduces to a ratio of volumes
\begin{equation}
    P(\boldsymbol x_{k+1} \notin \Omega_j | \boldsymbol x_k\in \Omega_j, \Delta\boldsymbol x_k) = \frac{V(\Omega_j\setminus(\Omega_j-\Delta\boldsymbol x_k))}{V(\Omega_j)}.
    \label{eq:conditional_escape_prob_step2}
\end{equation}

In conclusion, if Assumptions \ref{ass: uniform_x} and \ref{ass: ind_step} hold, then the conditional escape probability for a given $\Delta\boldsymbol x_k$ corresponds to the proportion of $\Omega_j$ that is not in $\Omega_j-\Delta\boldsymbol x_k$. This corresponds with the hatched area in Figure~\ref{fig: translation}. Indeed, this is the part of $\Omega_j$ which exits the cell when translated with $\Delta\boldsymbol x_k$.

\begin{remark}
    In Appendix~\ref{app: transition}, we extend this approach for computing escape probabilities to the computation of transition probabilities between two elements. The general approach is set out. A specific formula is computed for the one-dimensional case.
\end{remark}

\subsection{Local Coordinate Transform}\label{sec: local_coords}
\begin{figure}
    \centering
    \includegraphics[width=0.45\linewidth]{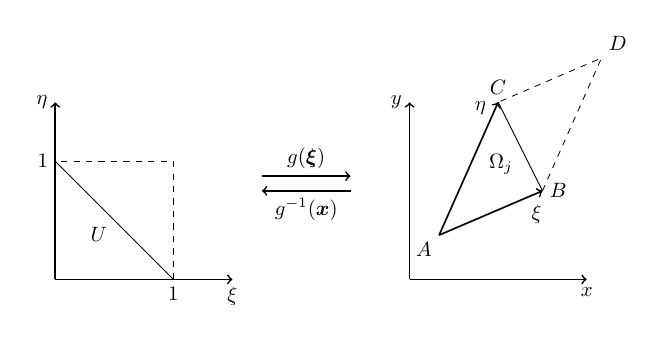}\qquad
    \includegraphics[width=0.45\linewidth]{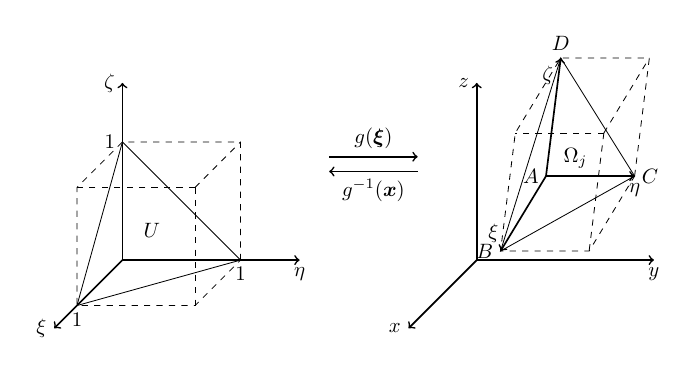}
    \caption{Transformation from local to global coordinates for triangles and parallelograms in two dimensions, and tetrahedra and parallelepipeds in three dimensions.}
    \label{fig: loc-to-glo}
\end{figure}

With the goal in mind of computing escape probabilities for common mesh elements (i.e., arbitrary lines, triangles, parallelograms, parallelepipeds, and tetrahedra), we introduce a transformation to local coordinates for each of these geometries. While the original escape probability problem is stated in a \textit{global coordinate system} $(x, y, z, \hdots)$, it will prove useful to transform to a cell's \textit{local coordinate system} $(\xi, \eta, \zeta, \hdots)$. We will exploit the fact that, in their respective local coordinate systems, all cells of the same type are identical to a common reference unit cell $U$. In the two-dimensional case, all parallelograms are identical to the unit square $S_2$, and all triangles are identical to the unit triangle $T_2$ (spanning $(0, 0)$, $(1, 0)$, and $(0, 1)$). In the three-dimensional case, similarly, we have that all parallelepipeds are identical the unit cube $C_3$, and all tetrahedrons to the unit tetrahedron $T_3$ (spanning $(0, 0, 0)$, $(1, 0, 0)$, $(0, 1, 0)$, and $(0, 0, 1)$). Figure~\ref{fig: loc-to-glo} shows these local coordinate systems for the different types of common mesh elements considered in this paper.

For these mesh elements, a point $\boldsymbol \xi$ in local coordinates is mapped to a point $\boldsymbol x$ in global coordinates by an affine transformation,
\begin{equation}
    \boldsymbol x = g(\boldsymbol \xi) = A\boldsymbol \xi+ \boldsymbol b,
    \label{eq:affine_transform}
\end{equation}
with full-rank $A\in\mathbb R^{n\times n}$ and $\boldsymbol b\in\mathbb R^{n\times 1}$, which depend on the specific geometry of $\Omega_j$. Because $A$ is full-rank, the inverse transformation $g^{-1}(\boldsymbol x) = A^{-1}(\boldsymbol x - \boldsymbol b)$ exists and is affine as well. It maps from global to local coordinates.

We want $g^{-1}(\boldsymbol x)$ to map the cell $\Omega_j$ to its reference unit cell $U$. That is, we choose $A$ and $\boldsymbol b$ such that
\begin{equation}
    \begin{split}
        g^{-1}(\Omega_j) &= U\\
        \Rightarrow g^{-1}(\Omega_j-\Delta\boldsymbol x_k) &= U - A^{-1}\Delta\boldsymbol x_k = U-\Delta\boldsymbol\xi_k.
    \end{split}\label{eq:prop1}
\end{equation}
We call $\Delta \boldsymbol \xi_k = A^{-1}\Delta \boldsymbol x_k$ the step in local coordinates corresponding to the step $\Delta \boldsymbol x_k$ in the global coordinate system.
Under affine transformations, ratios of volumes are preserved:
\begin{equation}
    \frac{V(A)}{V(B)} = \frac{V(g(A))}{V(g(B))} = \frac{V(g^{-1}(A))}{V(g^{-1}(B))}.
    \label{eq:prop2}
\end{equation}

Applying $g^{-1}(\boldsymbol x)$ to the numerator and denominator of~\eqref{eq:conditional_escape_prob_step2} and using~\eqref{eq:prop1} and~\eqref{eq:prop2} now yields
\begin{equation}
    \begin{split}
        P(\boldsymbol x_{k+1} \notin \Omega_j | \boldsymbol x_k\in \Omega_j, \Delta\boldsymbol x_k) &= \frac{V(\Omega_j\setminus(\Omega_j-\Delta\boldsymbol x_k))}{V(\Omega_j)}\\
        &= \frac{V(g^{-1}(\Omega_j\setminus(\Omega_j-\Delta\boldsymbol x_k)))}{V(g^{-1}(\Omega_j))}\\
        &= \frac{V(U\setminus(U-\Delta\boldsymbol\xi_k))}{V(U)}\\
        &= 1 - \frac{V(U\cap(U-\Delta\boldsymbol\xi_k))}{V(U)}.
    \end{split}
    \label{eq:conditional_escape_prob_step3}
\end{equation}
where we used that $g^{-1}(A\setminus B) = g^{-1}(A)\setminus g^{-1}(B)$, because $g^{-1}(\boldsymbol x)$ is injective. Note that the equalities in~\eqref{eq:conditional_escape_prob_step3} imply that
\begin{equation}
P(\boldsymbol x_{k+1} \notin \Omega_j | \boldsymbol x_k\in \Omega_j, \Delta\boldsymbol x_k) = P(\boldsymbol \xi_{k+1} \notin U | \boldsymbol \xi_k \in U, \Delta\boldsymbol \xi_k),
\label{eq:equivalence_localglobal}
\end{equation}

i.e., we have reduced the problem of determining the conditional escape probability for arbitrary geometries $\Omega_j$ to computing it only for the reference unit cells $U$. In Section \ref{sec:cond_escp_prob}, we solve the conditional escape probability problem for the line, unit triangle, unit square, unit tetrahedron, and
unit cube and specify the appropriate affine transformations required to generalize to any triangle, parallelogram, tetrahedron or parallelepiped.

\subsection{Integral Formulation in Local Coordinates}
While expression~\eqref{eq:conditional_escape_prob_step3} can be inserted in the integral formulation~\eqref{eq:integral_formulation} as is, it will prove advantageous to solve the entire integral in local coordinates. We use the change-of-variables formula~\cite{ben-israel_change--variables_1999} to transform to $\Delta\boldsymbol \xi = A^{-1}\Delta\boldsymbol x$
\begin{equation}
    \begin{split}
        P(\boldsymbol x_{k+1} \notin \Omega_j | \boldsymbol x_k \in \Omega_j) &= \int\limits_{\mathbb R^n} P(\boldsymbol x_{k+1} \notin \Omega_j | \boldsymbol x_k \in \Omega_j, \Delta\boldsymbol x)p(\Delta\boldsymbol x)\;\mathrm d(\Delta\boldsymbol x)\\
        &= \int\limits_{\mathbb R^n} P(\boldsymbol \xi_{k+1} \notin U | \boldsymbol \xi_k \in U, \Delta\boldsymbol \xi)p(A\Delta\boldsymbol \xi)|\det(A)|\;\mathrm d(\Delta\boldsymbol \xi).
    \end{split}
    \label{eq:integral_formulation_local_coordinates}
\end{equation}
The main advantage of the integral formulation in local coordinates over the one in global coordinates is the fact that the conditional escape probability expressions in Section~\ref{sec:cond_escp_prob} imply a subdivision of the integration domain into subdomains to avoid discontinuities in the integrand. This subdivision is the most natural in local coordinates. Additionally, in Section~\ref{sec:cond_escp_prob}, the expressions for the conditional escape probabilities are given in local coordinates, which implies that using them in~\eqref{eq:integral_formulation} would require solving a system $\Delta\boldsymbol \xi = A^{-1}\Delta\boldsymbol x$ for each evaluation of the integrand. The integral formulation in local coordinates does not require solving such systems.
Equation~\eqref{eq:integral_formulation_local_coordinates}, with the integrand in local coordinates, and the subdivision of the integration domain is implemented in the code accompanying this paper~\cite{code}.



\section{Conditional Escape Probability for Common Mesh Elements}
\label{sec:cond_escp_prob}
In Section \ref{sec:general_approach}, we explained our general approach to the computation of cell escape probabilities. Under Assumptions \ref{ass: uniform_x} and \ref{ass: ind_step}, we simplified the computation of the conditional escape probability, one of the two components in the integral formulation~\eqref{eq:integral_formulation}. By introducing a transformation to local coordinates, we reduced the general conditional escape probability problem to solving it for the specific reference unit cells (i.e. unit triangles, squares, tetrahedra, and cubes). In this section, we calculate conditional escape probabilities for these specific elements based on equation~\eqref{eq:conditional_escape_prob_step3}. For visualisation of the two-dimensional triangular and three-dimensional tetrahedral elements, several supporting GeoGebra tutorials are available at~\cite{code}.

\subsection{One-Dimensional Linear Mesh Elements}
\begin{figure}[H]
    \centering
    \includegraphics[width=0.4\linewidth]{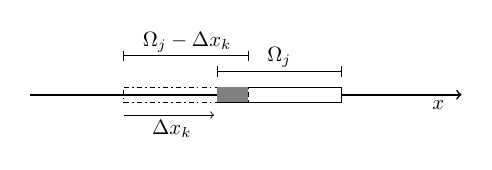}
    \caption{One-dimensional cell $\Omega_j$ of length $l$ and its translation $\Omega_j-\Delta x_k$. Under Assumptions \ref{ass: uniform_x} and \ref{ass: ind_step}, the conditional escape probability for a given $\Delta x_k$, corresponds to the proportion of $\Omega_j$ which is hatched.}
    \label{fig: 1d-translation}
\end{figure}

The one-dimensional case can be solved without transforming to local coordinates. Say $\Omega_j$ is a one-dimensional line segment of length $l$, hence $V(\Omega_j) = l$. For $\Delta x_k \in\mathbb R$, we have
\begin{equation}
    V(\Omega_j\cap(\Omega_j-\Delta x_k)) = \begin{cases}
        l-|\Delta x_k| & \mathrm{if} \quad |\Delta x_k| \leq l \\
        0              & \mathrm{otherwise}.
    \end{cases}
\end{equation}
We can compute the conditional escape probability via~\eqref{eq:conditional_escape_prob_step3} as
\begin{equation}
    P(x_{k+1}\notin \Omega_j | x_k\in \Omega_j, \Delta x_k) = \begin{cases}
        \frac{|\Delta x_k|}{l} & \mathrm{if} \quad |\Delta x_k| \leq l \\
        1                      & \mathrm{otherwise}.
    \end{cases}
\end{equation}
This is visualized in Figure~\ref{fig: 1d-translation}, which is a one-dimensional analogue to Figure~\ref{fig: translation}.

\subsection{Common Two-Dimensional Mesh Elements}
\subsubsection{Triangular Cells}
To calculate the conditional escape probability for an arbitrary triangular cell with vertices $A(x_A,y_A)$, $B(x_B,y_B)$, and $C(x_C,y_C)$, we make use of the affine transformation~\eqref{eq:affine_transform} with
\begin{equation}
    A = \begin{bmatrix}
        x_B-x_A & x_C-x_A \\
        y_B-y_A & y_C-y_A
    \end{bmatrix}\qquad\mathrm{and}\qquad \boldsymbol b = \begin{bmatrix}
        x_A \\
        y_A
    \end{bmatrix},
    \label{eq:affine_triangles}
\end{equation}
which maps the points $(0,0)$, $(0,1)$, and $(1,0)$ to $A$, $B$, and $C$ respectively.
As required, the unit triangle $T_2$ is mapped to $\Omega_j$:
\begin{equation}
    g(T_2) = \Omega_j.
\end{equation}

For an arbitrary step $\Delta\boldsymbol\xi = \begin{bmatrix}
        \Delta\xi \\\Delta\eta
    \end{bmatrix}$ in the local coordinate system, we have
\begin{equation}
    V(T_2\cap(T_2-\Delta\boldsymbol\xi)) = \begin{cases}
        \frac{1}{2}(1 - |\Delta \xi| - |\Delta \eta|)^2 & \mathrm{if} \quad 0\leq\Delta\xi\Delta\eta\ \mathrm{and}\ |\Delta\xi + \Delta\eta| \leq 1  \\
        \frac{1}{2}(1 - |\Delta \xi|)^2                 & \mathrm{if} \quad \Delta\eta(\Delta\xi+\Delta\eta) < 0\ \mathrm{and}\ |\Delta\xi| \leq 1   \\
        \frac{1}{2}(1 - |\Delta \eta|)^2                & \mathrm{if} \quad \Delta\xi(\Delta\xi + \Delta\eta) < 0\ \mathrm{and}\ |\Delta\eta| \leq 1 \\
        0                                               & \mathrm{otherwise}.
    \end{cases}
\end{equation}
The GeoGebra files at~\cite{code} serve to motivate these formulas. We also have
\begin{equation}
    V(T_2) = \frac{1}{2}.
\end{equation}

By substituting $\begin{bmatrix}
        \Delta\xi \\\Delta\eta
    \end{bmatrix} = A^{-1}\Delta\boldsymbol x_k$, we can compute the conditional escape probability via~\eqref{eq:conditional_escape_prob_step3} as
\begin{equation}
    P(\boldsymbol x_{k+1}\notin \Omega_j | \boldsymbol x_k\in \Omega_j, \Delta\boldsymbol x_k) = \begin{cases}
        1 - (1 - |\Delta \xi| - |\Delta \eta|)^2 & \mathrm{if} \quad 0\leq\Delta\xi\Delta\eta\ \mathrm{and}\ |\Delta\xi + \Delta\eta| \leq 1  \\
        1 - (1 - |\Delta \xi|)^2                 & \mathrm{if} \quad \Delta\eta(\Delta\xi+\Delta\eta) < 0\ \mathrm{and}\ |\Delta\xi| \leq 1   \\
        1 - (1 - |\Delta \eta|)^2                & \mathrm{if} \quad \Delta\xi(\Delta\xi + \Delta\eta) < 0\ \mathrm{and}\ |\Delta\eta| \leq 1 \\
        1                                        & \mathrm{otherwise}.
    \end{cases}
\end{equation}

\subsubsection{Parallelogrammatic Cells}
To calculate the conditional escape probability for parallelogrammatic cells $\Omega_j$, the same affine transformation as for triangular cells, defined by~\eqref{eq:affine_triangles}, can be used. If vertices $A$, $B$, and $C$ are chosen as in Figure \ref{fig: loc-to-glo}, this transformation maps the unit square $S_2$ to $\Omega_j$:

\begin{equation}
    g(S_2) = \Omega_j.
\end{equation}

For an arbitrary step $\Delta\boldsymbol\xi = \begin{bmatrix}
        \Delta\xi \\\Delta\eta
    \end{bmatrix}$ in the local coordinates system, we have
\begin{equation}
    V(S_2\cap(S_2-\Delta\boldsymbol\xi)) = \begin{cases}
        (1 - |\Delta\xi|)(1 - |\Delta\eta|) & \mathrm{if} \quad \max(|\Delta\xi|, |\Delta\eta|)\leq 1 \\
        0                                   & \mathrm{otherwise}.
    \end{cases}
\end{equation}

By substituting $\begin{bmatrix}
        \Delta\xi \\\Delta\eta
    \end{bmatrix} = A^{-1}\Delta\boldsymbol x_k$, we can compute the conditional escape probability via~\eqref{eq:conditional_escape_prob_step3} as
\begin{equation}
    P(\boldsymbol x_{k+1}\notin \Omega_j | \boldsymbol x_k\in \Omega_j,\Delta\boldsymbol x_k) = \begin{cases}
        1 - (1 - |\Delta\xi|)(1 - |\Delta\eta|) & \mathrm{if} \quad \max(|\Delta\xi|, |\Delta\eta|) \leq 1 \\
        1                                       & \mathrm{otherwise}.
    \end{cases}
\end{equation}

\subsection{Common Three-Dimensional Mesh Elements}
\subsubsection{Tetrahedral Cells}
Similar as for two-dimensional mesh elements, we calculate the conditional escape probability for tetrahedral cells using an affine transformation~\eqref{eq:affine_transform} with
\begin{equation}
    A = \begin{bmatrix}
        x_B-x_A & x_C-x_A & x_D-x_A \\
        y_B-y_A & y_C-y_A & y_D-y_A \\
        z_B-z_A & z_C-z_A & z_D-z_A
    \end{bmatrix}\qquad\mathrm{and}\qquad \boldsymbol b = \begin{bmatrix}
        x_A \\
        y_A \\
        z_A
    \end{bmatrix},
    \label{eq:affine_tetrahedral}
\end{equation}
which maps the points $(0, 0, 0)$, $(1, 0, 0)$, $(0, 1, 0)$, and $(0, 0, 1)$ to some $A(x_A, y_A, z_A)$, $B(x_B, y_B, z_B)$, $C(x_C, y_C, z_C)$, and $D(x_D, y_D, z_D)$ respectively. Consequently, the transformation maps the unit tetrahedron $T_3$ to an arbitrary tetrahedral cell $\Omega_j$ with vertices $A$, $B$, $C$, and $D$:
\begin{equation}
    g(T_3) = \Omega_j.
\end{equation}

For an arbitrary step $\Delta\boldsymbol\xi = \begin{bmatrix}
        \Delta\xi \\\Delta\eta\\\Delta\zeta
    \end{bmatrix}$ in the local coordinate system, we have
\begin{equation}
    \begin{split}
        & V(T_3\cap(T_3-\Delta\boldsymbol\xi)) = \\
        & \hspace{4mm} \begin{cases}
            \frac{1}{6}(1 - |\Delta \xi| - |\Delta \eta| - |\Delta\zeta|)^3 & \mathrm{if} \quad 0\leq\Delta\xi\Delta\eta, 0\leq\Delta\xi\Delta\zeta\ \mathrm{and}\ |\Delta\xi + \Delta\eta + \Delta\zeta| \leq 1                              \\
            \frac{1}{6}(1 - |\Delta \eta| - |\Delta\zeta|)^3                & \mathrm{if} \quad \Delta\xi\Delta\eta<0, \Delta\xi\Delta\zeta<0, \Delta\xi(\Delta\xi+\Delta\eta+\Delta\zeta)<0\ \mathrm{and}\ |\Delta\eta+\Delta\zeta| \leq 1   \\
            \frac{1}{6}(1 - |\Delta\xi| - |\Delta \eta|)^3                  & \mathrm{if} \quad \Delta\xi\Delta\eta<0, \Delta\eta\Delta\zeta<0, \Delta\eta(\Delta\xi+\Delta\eta+\Delta\zeta)<0\ \mathrm{and}\ |\Delta\xi+\Delta\zeta| \leq 1  \\
            \frac{1}{6}(1 - |\Delta\xi| - |\Delta \zeta|)^3                 & \mathrm{if} \quad \Delta\xi\Delta\zeta<0, \Delta\eta\Delta\zeta<0, \Delta\zeta(\Delta\xi+\Delta\eta+\Delta\zeta)<0\ \mathrm{and}\ |\Delta\xi+\Delta\eta| \leq 1 \\
            \frac{1}{6}(1 - |\Delta \xi|)^3                                 & \mathrm{if} \quad \Delta\eta(\Delta\xi+\Delta\eta+\Delta\zeta)<0, \Delta\zeta(\Delta\xi+\Delta\eta+\Delta\zeta)<0\ \mathrm{and}\ |\Delta\xi| \leq 1             \\
            \frac{1}{6}(1 - |\Delta \eta|)^3                                & \mathrm{if} \quad \Delta\xi(\Delta\xi+\Delta\eta+\Delta\zeta)<0, \Delta\zeta(\Delta\xi+\Delta\eta+\Delta\zeta)<0\ \mathrm{and}\ |\Delta\eta| \leq 1             \\
            \frac{1}{6}(1 - |\Delta \zeta|)^3                               & \mathrm{if} \quad \Delta\xi(\Delta\xi+\Delta\eta+\Delta\zeta)<0, \Delta\eta(\Delta\xi+\Delta\eta+\Delta\zeta)<0\ \mathrm{and}\ |\Delta\zeta| \leq 1             \\
            0                                                               & \mathrm{otherwise}.
        \end{cases}
    \end{split}
\end{equation}
As for the triangular cells, the GeoGebra files available at~\cite{code} motivate these formulas. We also have
\begin{equation}
    V(T_3) = \frac{1}{6}.
\end{equation}

By substituting $\begin{bmatrix}
        \Delta\xi \\\Delta\eta\\\Delta\zeta
    \end{bmatrix} = A^{-1}\Delta\boldsymbol x_k$, we can compute the conditional escape probability via~\eqref{eq:conditional_escape_prob_step3} as
\begin{equation}
    \begin{split}
        &P(\boldsymbol x_{k+1}\notin \Omega_j | \boldsymbol x_k\in \Omega_j, \Delta\boldsymbol x_k) = \\
        & \hspace{4mm} \begin{cases}
            1 - (1 - |\Delta \xi| - |\Delta \eta| - |\Delta\zeta|)^3 & \mathrm{if} \quad 0\leq\Delta\xi\Delta\eta, 0\leq\Delta\xi\Delta\zeta\ \mathrm{and}\ |\Delta\xi + \Delta\eta + \Delta\zeta| \leq 1                              \\
            1 - (1 - |\Delta \eta| - |\Delta\zeta|)^3                & \mathrm{if} \quad \Delta\xi\Delta\eta<0, \Delta\xi\Delta\zeta<0, \Delta\xi(\Delta\xi+\Delta\eta+\Delta\zeta)<0\ \mathrm{and}\ |\Delta\eta+\Delta\zeta| \leq 1   \\
            1 - (1 - |\Delta\xi| - |\Delta \zeta|)^3                 & \mathrm{if} \quad \Delta\xi\Delta\eta<0, \Delta\eta\Delta\zeta<0, \Delta\eta(\Delta\xi+\Delta\eta+\Delta\zeta)<0\ \mathrm{and}\ |\Delta\xi+\Delta\zeta| \leq 1  \\
            1 - (1 - |\Delta\xi| - |\Delta \eta|)^3                  & \mathrm{if} \quad \Delta\xi\Delta\zeta<0, \Delta\eta\Delta\zeta<0, \Delta\zeta(\Delta\xi+\Delta\eta+\Delta\zeta)<0\ \mathrm{and}\ |\Delta\xi+\Delta\eta| \leq 1 \\
            1 - (1 - |\Delta \xi|)^3                                 & \mathrm{if} \quad \Delta\eta(\Delta\xi+\Delta\eta+\Delta\zeta)<0, \Delta\zeta(\Delta\xi+\Delta\eta+\Delta\zeta)<0\ \mathrm{and}\ |\Delta\xi| \leq 1             \\
            1 - (1 - |\Delta \eta|)^3                                & \mathrm{if} \quad \Delta\xi(\Delta\xi+\Delta\eta+\Delta\zeta)<0, \Delta\zeta(\Delta\xi+\Delta\eta+\Delta\zeta)<0\ \mathrm{and}\ |\Delta\eta| \leq 1             \\
            1 - (1 - |\Delta \zeta|)^3                               & \mathrm{if} \quad \Delta\xi(\Delta\xi+\Delta\eta+\Delta\zeta)<0, \Delta\eta(\Delta\xi+\Delta\eta+\Delta\zeta)<0\ \mathrm{and}\ |\Delta\zeta| \leq 1             \\
            1                                                        & \mathrm{otherwise}.
        \end{cases}
    \end{split}
\end{equation}

\subsubsection{Parallelepipedal Cells}
Finally, for parallelepipedal cells $\Omega_j$, we use the same affine transformation as for tetrahedral cells, defined by~\eqref{eq:affine_tetrahedral}, to calculate the conditional escape probability. If vertices $A$, $B$, $C$, and $D$ are chosen as shown in Figure \ref{fig: loc-to-glo}, it maps the unit cube $C_3$ to $\Omega_j$:
\begin{equation}
    g(C_3) = \Omega_j.
\end{equation}

For an arbitrary step $\Delta\boldsymbol\xi = \begin{bmatrix}
        \Delta\xi \\\Delta\eta\\\Delta\zeta
    \end{bmatrix}$ in the local coordinate system, we have
\begin{equation}
    V(C_3\cap(C_3-\Delta\boldsymbol\xi)) = \begin{cases}
        (1 - |\Delta\xi|)(1 - |\Delta\eta|)(1 - |\Delta\zeta|) & \mathrm{if} \quad \max(|\Delta\xi|, |\Delta\eta|, |\Delta\zeta|)\leq 1 \\
        0                                                      & \mathrm{otherwise}.
    \end{cases}
\end{equation}

By substituting $\begin{bmatrix}
        \Delta\xi \\\Delta\eta\\\Delta\zeta
    \end{bmatrix} = A^{-1}\Delta\boldsymbol x_k$, we can compute the conditional escape probability via~\eqref{eq:conditional_escape_prob_step3} as
\begin{equation}
    P(\boldsymbol x_{k+1}\notin \Omega_j | \boldsymbol x_k\in \Omega_j, \Delta\boldsymbol x_k) = \begin{cases}
        1 - (1 - |\Delta\xi|)(1 - |\Delta\eta|)(1 - |\Delta\zeta|) & \mathrm{if} \quad \max(|\Delta\xi|, |\Delta\eta|, |\Delta\zeta|)\leq 1 \\
        1                                                          & \mathrm{otherwise}.
    \end{cases}
\end{equation}

Substituting these expressions for the conditional escape probability and the appropriate transformation matrix $A$ into~\eqref{eq:integral_formulation_local_coordinates} and evaluating the integral yields the sought-after escape probability.

\section{Monte Carlo Algorithm}
\label{sec: mca}
In some cases, solving the cell escape probability integral formulation~\eqref{eq:integral_formulation_local_coordinates} in a deterministic way becomes prohibitive. This happens, for instance, when either Assumption \ref{ass: uniform_x} or \ref{ass: ind_step} is not met, when the integrals are high-dimensional, or when the step probability density $p(\Delta\boldsymbol x | \boldsymbol x_k,\boldsymbol s_k,t_k)$ is a function that is difficult to integrate. In these cases, a Monte Carlo algorithm provides a robust alternative for calculating the escape probability at the cost of a statistical error. In Section~\ref{sec:numerical_results}, we use this Monte Carlo algorithm to verify the correctness of our deterministic solver for the cell escape probability integral formulation~\eqref{eq:integral_formulation_local_coordinates}.

The Monte Carlo algorithm, given in Algorithm~\ref{algorithm}, obtains an estimate of the cell escape probability using three steps. First, $N$ particle positions $\{\boldsymbol x_{k,n}\}_{n=1}^N$ are sampled from $p(\boldsymbol x|\boldsymbol x\in\Omega_j, t_k)$. This probability density is zero outside $\Omega_j$. Inside $\Omega_j$, it is equal to $p(\boldsymbol x|t_k)$ up to a scaling factor for normalization. Second, each particle independently executes one step of the Markov process~\eqref{eq:markov_process}. Third, the escape probability is estimated as the proportion of particles that exited $\Omega_j$ during this step.

The resulting cell escape probability estimate is quantized, but unbiased. The statistical error on the estimate corresponds to the standard deviation of a binomial experiment with $N$ independent trials, where for each successful escape, a contribution of $\frac{1}{N}$ is counted and where the probability for success equals the cell escape probability: 
\begin{equation}
    |E_{s}| = \sqrt{\frac{1}{N} \cdot P(\boldsymbol x_{k+1}\notin \Omega_j | \boldsymbol x_k\in \Omega_j) \cdot (1 - P(\boldsymbol x_{k+1}\notin \Omega_j | \boldsymbol x_k\in \Omega_j))}.
    \label{eq:stat_error}
\end{equation}
An implementation of the Monte Carlo algorithm with documentation providing additional explanation on the different steps of the algorithm is provided in the code accompanying this paper~\cite{code}.

\begin{algorithm}[H]
    \begin{algorithmic}
        \State Given a number of particles $N$, probability density functions $p(\boldsymbol x|t_k)$ and $p(\Delta\boldsymbol x|\boldsymbol x_k, \boldsymbol s_k, t_k)$, and the geometry of the cell $\Omega_j$ from which we want to calculate the escape probability.
        \State
        \For{n=1,\dots, N}
        \State $\boldsymbol x_{k,n} \sim p(\boldsymbol x|\boldsymbol x \in\Omega_j, t_k)$
        \Comment{Sample from $p(\boldsymbol x|t_k)$ ensuring that $\boldsymbol x_{k,n}\in\Omega_j$}
        \State $\Delta\boldsymbol x_{k,n} \sim p(\Delta\boldsymbol x|\boldsymbol x_{k,n}, \boldsymbol s_{k,n}, t_k)$
        \State $\boldsymbol x_{k+1,n} \gets \boldsymbol x_{k,n} + \Delta\boldsymbol x_{k,n}$
        \Comment{Simulate one step of the Markov process}
        \EndFor
        \State \textbf{return} $\frac{\#\{\boldsymbol x_{k+1,n}\notin\Omega_j\}}{N}$
        \Comment{Return the number of particles outside $\Omega_j$ divided by the total number of particles}
    \end{algorithmic}

    \caption{Monte Carlo algorithm for estimating the cell escape probability.}
    \label{algorithm}
\end{algorithm}

\section{Numerical Results}
\label{sec:numerical_results}
In this section, we verify the correctness of the analysis and implementation of our deterministic solver and the Monte Carlo algorithm in the code accompanying this paper~\cite{code}.

We consider a particle undergoing the $n$-dimensional Wiener process described in Example \ref{ex:wiener} in Section~\ref{sec:introduction}, over a time interval $\Delta t$. The initial position $\boldsymbol x_0$ of the particle is random and uniformly distributed in a cell $\Omega_j$. For this Wiener process, the step is sampled from an isotropic normal distribution with zero mean step size and variance $\Delta t$ in all directions:
\begin{align}
    \boldsymbol x_0          & \sim p(\boldsymbol x |t=0) = \begin{cases}
                                                                \frac{1}{V(\Omega_j)} \qquad & \mathrm{if} \quad \boldsymbol x \in \Omega_j \\
                                                                0                            & \mathrm{otherwise},
                                                            \end{cases} \\
    \Delta \boldsymbol x_{0} & \sim p(\Delta\boldsymbol x) = N(0,\Delta t I_n).
\end{align}

The goal is to compute the probability that the particle exits $\Omega_j$ within the time interval. We do this in two ways: we calculate the escape probability deterministically by using the standard MATLAB routines \texttt{integral()}, \texttt{integral2()}, and \texttt{integral3()} to evaluate the integral formulation~\eqref{eq:integral_formulation_local_coordinates}, and stochastically by using a sequential implementation of the Monte Carlo algorithm explained in Section~\ref{sec: mca}. We verify our results for one instance of each of the common mesh elements covered in this paper - i.e., lines, triangles, parallelograms, parallelepipeds, and tetrahedra. The specific geometries used in the experiments are summarized in Table \ref{tab: geometries}.

\begin{table}[H]
    \begin{adjustbox}{center}
        \begin{tabular}{r|llllllll}
            \textbf{Geometry}   & \textbf{Vertices}                                                                                            \\
            \hline
            Line (1D)           & $A(0)$            & $B(2)$     &            &            &            &            &            &            \\
            Triangle (2D)       & $A(0, 0)$         & $B(2, 0)$  & $C(3, 2)$  &            &            &            &            &            \\
            Parallelogram (2D)  & $A(0, 0)$         & $B(2, 0)$  & $C(1, 2)$  & $D(3, 2)$  &            &            &            &            \\
            Tetrahedron (3D)    & $A(0,0,0)$        & $B(2,0,0)$ & $C(3,2,0)$ & $D(1,1,1)$ &            &            &            &            \\
            Parallelepiped (3D) & $A(0,0,0)$        & $B(2,0,0)$ & $C(1,2,1)$ & $D(0,0,2)$ & $E(3,2,1)$ & $F(2,0,2)$ & $G(3,2,3)$ & $H(1,2,3)$
        \end{tabular}
    \end{adjustbox}
    \caption{Coordinates of the vertices for each instance of the cell types used in the numerical experiments.}
    \label{tab: geometries}
\end{table}

Table~\ref{table:results} shows the results of the numerical experiments for different values of $\Delta t$. For larger values of $\Delta t$, the particles are more likely to take larger steps $\Delta \boldsymbol x_0$, resulting in larger escape probabilities. The table shows the calculated escape probabilities, together with the computational time for the deterministic solver and the Monte Carlo algorithm with $N = 10^6$ particles.  The estimated error of the deterministic solver, the empirical statistical error based on 10 runs of the Monte Carlo algorithm, and the theoretical statistical error~\eqref{eq:stat_error} of the Monte Carlo algorithm are also displayed. The theoretical statistical error is based on the deterministic solver result for the escape probability. The difference in results between the deterministic solver and the Monte Carlo algorithm can be explained by the theoretical statistical error. In all cases, the deterministic solver outperforms the Monte Carlo algorithm both in runtime and accuracy. The Monte Carlo algorithm, however, remains a good tool to verify the implementation of the deterministic solver and can be used as a robust alternative for the deterministic solver for problems where the step probability density is more involved.

Note that both the deterministic solver and the Monte Carlo algorithm become more expensive as the dimension of the mesh element increases. For the Monte Carlo method this is due to the third step of the algorithm: verifying if the particle has escaped becomes more expensive in higher dimensions. Also note that the deterministic solver is more expensive for lower escape probabilities, because the adaptive integration algorithm then has to perform more refinement work.

\begin{table}[!ht]
    \centering
    \def\arraystretch{1.25}
    \begin{tabular}{|l|l|l|l|l|l|l|}
        \hline
        \textbf{1D linear element}             & $\Delta t = 10^{-2}$ & $\Delta t = 10^{-1}$ & $\Delta t = 10^{0}$ & $\Delta t = 10^{1}$ & $\Delta t = 10^{2}$ & $\Delta t = 10^{3}$ \\ \hline\hline
        Det. solver result                     & 0.0399               & 0.1262               & 0.3905              & 0.7558              & 0.9205              & 0.9748              \\ \hline
        MC algorithm result                    & 0.0398               & 0.1267               & 0.3901              & 0.7562              & 0.9211              & 0.9748              \\ \hline \hline
        Det. solver timing                     & 0.0510               & 0.0082               & 0.0080              & 0.0016              & 0.0044              & 0.0009              \\ \hline
        MC algorithm timing                    & 0.0795               & 0.0668               & 0.0727              & 0.0797              & 0.0801              & 0.0.0809            \\ \hline \hline
        Det. solver error estimate             & $1\cdot 10^{-6}$     & $1\cdot 10^{-6}$     & $1\cdot 10^{-6}$    & $1\cdot 10^{-6}$    & $1\cdot 10^{-6}$    & $1\cdot 10^{-6}$    \\ \hline
        MC algorithm error estimate            & $1.7 \cdot 10^{-4}$  & $2.9\cdot 10^{-4}$   & $2.6\cdot 10^{-4}$  & $3.1\cdot 10^{-4}$  & $2.4\cdot 10^{-4}$  & $1.4\cdot 10^{-4}$  \\ \hline
        Theoretical statistical error          & $2.0\cdot 10^{-4}$   & $3.3\cdot 10^{-4}$   & $4.9\cdot 10^{-4}$  & $4.3\cdot 10^{-4}$  & $2.7\cdot 10^{-4}$  & $1.6\cdot 10^{-4}$  \\ \hline \hline
        \textbf{2D triangular element }        & $\Delta t = 10^{-2}$ & $\Delta t = 10^{-1}$ & $\Delta t = 10^{0}$ & $\Delta t = 10^{1}$ & $\Delta t = 10^{2}$ & $\Delta t = 10^{3}$ \\ \hline\hline
        Det. solver result                     & 0.1478               & 0.4082               & 0.7957              & 0.9700              & 0.9968              & 0.9997              \\ \hline
        MC algorithm result                    & 0.1477               & 0.4079               & 0.7959              & 0.9700              & 0.9970              & 0.9997              \\ \hline \hline
        Det. solver timing                     & 0.1864               & 0.0481               & 0.0324              & 0.0328              & 0.0284              & 0.0269              \\ \hline
        MC algorithm timing                    & 13.8029              & 13.7501              & 13.7248             & 13.5106             & 13.1777             & 13.2370             \\ \hline \hline
        Det. solver error estimate             & $1\cdot 10^{-6}$     & $1\cdot 10^{-6}$     & $1\cdot 10^{-6}$    & $1\cdot 10^{-6}$    & $1\cdot 10^{-6}$    & $1\cdot 10^{-6}$    \\ \hline
        MC algorithm error estimate            & $3.9\cdot 10^{-4}$   & $8.2\cdot 10^{-4}$   & $5.0\cdot 10^{-4}$  & $1.1\cdot 10^{-4}$  & $7.9\cdot 10^{-5}$  & $1.5\cdot 10^{-5}$  \\ \hline
        Theoretical statistical error          & $3.5\cdot 10^{-4}$   & $4.9\cdot 10^{-4}$   & $4.0\cdot 10^{-4}$  & $1.7\cdot 10^{-4}$  & $5.6\cdot 10^{-5}$  & $1.8\cdot 10^{-5}$  \\ \hline \hline
        \textbf{2D parallelogrammatic element} & $\Delta t = 10^{-2}$ & $\Delta t = 10^{-1}$ & $\Delta t = 10^{0}$ & $\Delta t = 10^{1}$ & $\Delta t = 10^{2}$ & $\Delta t = 10^{3}$ \\ \hline\hline
        Det. solver result                     & 0.0825               & 0.2476               & 0.6394              & 0.9408              & 0.9937              & 0.9994              \\ \hline
        MC algorithm result                    & 0.0823               & 0.2481               & 0.6390              & 0.9413              & 0.9939              & 0.9994              \\ \hline \hline
        Det. solver timing                     & 0.0310               & 0.0205               & 0.0074              & 0.0075              & 0.0063              & 0.0050              \\ \hline
        MC algorithm timing                    & 6.9259               & 6.8452               & 6.6233              & 6.4372              & 6.2476              & 6.3855              \\ \hline \hline
        Det. solver error estimate             & $1\cdot 10^{-6}$     & $1\cdot 10^{-6}$     & $1\cdot 10^{-6}$    & $1\cdot 10^{-6}$    & $1\cdot 10^{-6}$    & $1\cdot 10^{-6}$    \\ \hline
        MC algorithm error estimate            & $1.8\cdot 10^{-4}$   & $4.2\cdot 10^{-4}$   & $4.3\cdot 10^{-4}$  & $3.0\cdot 10^{-4}$  & $9.4\cdot 10^{-5}$  & $3.0\cdot 10^{-5}$  \\ \hline
        Theoretical statistical error          & $2.8\cdot 10^{-4}$   & $4.3\cdot 10^{-4}$   & $4.8\cdot 10^{-4}$  & $2.4\cdot 10^{-4}$  & $7.9\cdot 10^{-5}$  & $2.5\cdot 10^{-5}$  \\ \hline \hline
        \textbf{3D tetrahedral element}        & $\Delta t = 10^{-2}$ & $\Delta t = 10^{-1}$ & $\Delta t = 10^{0}$ & $\Delta t = 10^{1}$ & $\Delta t = 10^{2}$ & $\Delta t = 10^{3}$ \\ \hline\hline
        Det. solver result                     & 0.3534               & 0.7433               & 0.9701              & 0.9987              & 1.0000              & 1.0000              \\ \hline
        MC algorithm result                    & 0.3530               & 0.7430               & 0.9700              & 0.9987              & 1.0000              & 1.0000              \\ \hline \hline
        Det. solver timing                     & 7.7701               & 3.7529               & 2.7734              & 2.5465              & 2.3105              & 1.9324              \\ \hline
        MC algorithm timing                    & 14.3715              & 14.1962              & 13.9057             & 13.6718             & 13.9680             & 15.2500             \\ \hline \hline
        Det. solver error estimate             & $1\cdot 10^{-6}$     & $1\cdot 10^{-6}$     & $1\cdot 10^{-6}$    & $1\cdot 10^{-6}$    & $1\cdot 10^{-6}$    & $1\cdot 10^{-6}$    \\ \hline
        MC algorithm error estimate            & $4.5\cdot 10^{-4}$   & $3.2\cdot 10^{-4}$   & $1.6\cdot 10^{-4}$  & $3.3\cdot 10^{-5}$  & $8.0\cdot 10^{-6}$  & $1.4\cdot 10^{-6}$  \\ \hline
        Theoretical statistical error          & $4.8\cdot 10^{-4}$   & $4.4\cdot 10^{-4}$   & $1.7\cdot 10^{-4}$  & $3.6\cdot 10^{-5}$  & $6.5\cdot 10^{-6}$  & $1.2\cdot 10^{-6}$  \\ \hline \hline
        \textbf{3D parallelepipedal element}   & $\Delta t = 10^{-2}$ & $\Delta t = 10^{-1}$ & $\Delta t = 10^{0}$ & $\Delta t = 10^{1}$ & $\Delta t = 10^{2}$ & $\Delta t = 10^{3}$ \\ \hline\hline
        Det. solver result                     & 0.1232               & 0.3519               & 0.7864              & 0.9856              & 0.9995              & 1.0000              \\ \hline
        MC algorithm result                    & 0.1233               & 0.3515               & 0.7859              & 0.9856              & 0.9995              & 1.0000              \\ \hline \hline
        Det. solver timing                     & 3.2163               & 1.0181               & 0.5466              & 0.3256              & 0.3461              & 0.3312              \\ \hline
        MC algorithm timing                    & 9.4680               & 9.4058               & 9.1933              & 8.8586              & 9.0153              & 8.8947              \\ \hline \hline
        Det. solver error estimate             & $1\cdot 10^{-6}$     & $1\cdot 10^{-6}$     & $1\cdot 10^{-6}$    & $1\cdot 10^{-6}$    & $1\cdot 10^{-6}$    & $1\cdot 10^{-6}$    \\ \hline
        MC algorithm error estimate            & $2.8\cdot 10^{-4}$   & $4.3\cdot 10^{-4}$   & $3.9\cdot 10^{-4}$  & $1.7\cdot 10^{-4}$  & $1.9\cdot 10^{-5}$  & $2.7\cdot 10^{-6}$  \\ \hline
        Theoretical statistical error          & $3.3\cdot 10^{-4}$   & $4.8\cdot 10^{-4}$   & $4.1\cdot 10^{-4}$  & $1.2\cdot 10^{-4}$  & $2.2\cdot 10^{-5}$  & $4.0\cdot 10^{-6}$  \\ \hline
    \end{tabular}
    \caption{Escape probabilities (result), timings (in seconds), and absolute error estimates for the deterministic solver (Det. solver) and the Monte Carlo algorithm (MC algorithm) for the different mesh elements covered in this paper. For the Monte Carlo algorithm, the theoretical statistical error~\eqref{eq:stat_error} using the escape probability calculated by the deterministic solver is also tabulated.}
    \label{table:results}
\end{table}

\section{Conclusion}
\label{sec: conclusion}
When executing a Markov process on a grid, the cell escape probability is of interest for applications such as statistical error prediction and the construction of diffusion Monte Carlo-like algorithms. The cell escape probabilities can be calculated by deterministically solving an integral formulation, or by using a stochastic Monte Carlo algorithm. Deterministically solving the integral formulation typically provides accurate solutions, but for difficult and high-dimensional integrals, the Monte Carlo algorithm provides a robust alternative at the cost of a statistical error. In this paper, the integral formulation and a Monte Carlo algorithm are formulated for common mesh elements: linear elements in one dimension, triangles and parallelograms in two dimensions, and tetrahedrons and parallelepipeds in three dimensions. The numerical experiments verify the correctness of the implemented deterministic solver and the Monte Carlo algorithm. One possible direction for future work is to try to include domain boundary interactions in the integral formulation for the cell escape probability. A second direction is to extend framework for calculating escape probabilities to the calculation of transition probabilities in higher dimensions.
\section*{Acknowledgement}
Vince Maes is an SB PhD fellow of the Research Foundation Flanders (FWO), funded by grant 1S64723N.

\newpage
\appendix

\section{Cell transition probabilities}
\label{app: transition}
The general approach for calculating escape probabilities from a grid cell can be extended to the calculation of transition probabilities between two grid cells $\Omega_1$ and $\Omega_2$. Define the \textit{transition probability} as $P(\boldsymbol x_{k+1} \in \Omega_2|\boldsymbol x_k \in\Omega_1)$. Similar to Section~\ref{sec: integral}, if Assumptions \ref{ass: uniform_x} and \ref{ass: ind_step} hold for all $\boldsymbol x\in\Omega_1$, this transition probability can be expressed as an integral over all possible values for the step $\Delta\boldsymbol x$:
\begin{equation}
    P(\boldsymbol x_{k+1} \in \Omega_2 | \boldsymbol x_k \in \Omega_1) = \int\limits_{\mathbb R^n} P(\boldsymbol x_{k+1} \in \Omega_2 | \boldsymbol x_k \in \Omega_1, \Delta\boldsymbol x)  p(\Delta\boldsymbol x)\;\mathrm d(\Delta\boldsymbol x).
    \label{eq:transition-prob}
\end{equation}
The step probability density $p(\Delta\boldsymbol x)$ again follows from the dynamics at hand. We call $P(\boldsymbol x_{k+1} \in \Omega_2 | \boldsymbol x_k \in \Omega_1, \Delta\boldsymbol x_k)$ the \textit{conditional transition probability} between $\Omega_1$ and $\Omega_2$, for a given step $\Delta\boldsymbol x_k$.
\begin{equation}
    \begin{split}
        P(\boldsymbol x_{k+1} \in \Omega_2 | \boldsymbol x_k \in \Omega_1, \Delta\boldsymbol x_k) &= \frac{P(\boldsymbol x_k\in \Omega_1\ \mathrm{and}\ \boldsymbol x_{k+1}\in\Omega_2|\Delta\boldsymbol x_k)}{P(\boldsymbol x_k\in\Omega_1 |  \Delta\boldsymbol x_k)}\\
        &= \frac{P(\boldsymbol x_k\in \Omega_1\ \mathrm{and}\ \boldsymbol x_k\in(\Omega_2-\Delta\boldsymbol x_k)| \Delta\boldsymbol x_k)}{P(\boldsymbol x_k\in\Omega_1 |  \Delta\boldsymbol x_k)}\\
        &= \frac{P(\boldsymbol x_k\in \Omega_1\cap(\Omega_2-\Delta\boldsymbol x_k)| \Delta\boldsymbol x_k)}{P(\boldsymbol x_k\in\Omega_1| \Delta\boldsymbol x_k)}.\\
    \end{split}
    \label{eq:conditional_transition_prob_step1}
\end{equation}
Under Assumption \ref{ass: uniform_x} and \ref{ass: ind_step}, the numerator and denominator in~\eqref{eq:conditional_transition_prob_step1} can respectively be computed as
\begin{equation}
    \begin{split}
        P(\boldsymbol x_k \in \Omega_1 \cap (\Omega_2-\Delta\boldsymbol x_k)| \Delta\boldsymbol x_k) & = \int\limits_{\mathclap{\Omega_1\cap(\Omega_2-\Delta\boldsymbol x_k)}}\frac{p(\boldsymbol x,\Delta\boldsymbol x_k)}{p(\Delta\boldsymbol x_k)}\;\mathrm d\boldsymbol x \\
        & = cV(\Omega_1 \cap (\Omega_2-\Delta\boldsymbol x_k))
    \end{split}
\end{equation}
and
\begin{equation}
    \begin{split}
        P(\boldsymbol x_k \in \Omega_1| \Delta\boldsymbol x_k) & = \int\limits_{\Omega_1}\frac{p(\boldsymbol x,\Delta\boldsymbol x_k)}{p(\Delta\boldsymbol x_k)}\;\mathrm d\boldsymbol x \\
        & = cV(\Omega_1).
    \end{split}
\end{equation}
We can simplify the conditional transition probability as follows:
\begin{equation}
    \begin{split}
        P(\boldsymbol x_{k+1} \in \Omega_2 | \boldsymbol x_k\in \Omega_1, \Delta\boldsymbol x_k) &= \frac{P(\boldsymbol x_k\in \Omega_1\cap(\Omega_2-\Delta\boldsymbol x_k)| \Delta\boldsymbol x_k)}{P(\boldsymbol x_k\in\Omega_1| \Delta\boldsymbol x_k)}\\
        &= \frac{V(\Omega_1\cap(\Omega_2-\Delta\boldsymbol x_k))}{V(\Omega_1)}.
    \end{split}
    \label{eq:conditional_transition_prob_step2}
\end{equation}

\begin{figure}[H]
    \centering
    \includegraphics[width=0.4\linewidth]{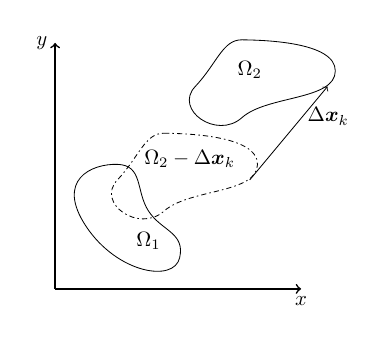}
    \caption{Cell $\Omega_1$, cell $\Omega_2$ and translation $\Omega_2-\Delta\boldsymbol x_k$. If Assumptions \ref{ass: uniform_x} and \ref{ass: ind_step} hold, then the conditional transition probability for a given $\Delta\boldsymbol x_k$ corresponds to the proportion of $\Omega_1$ that is also in $\Omega_2-\Delta\boldsymbol x_k$, which is hatched.}
    \label{fig: 2d-translation-transition}
\end{figure}

This gives us a general approach to compute conditional transition probabilities between two cells $\Omega_1$ and $\Omega_2$, under Assumptions~\ref{ass: uniform_x} and~\ref{ass: ind_step}. The conditional transition probability equals the proportion of $\Omega_1$ that is also in $\Omega_2-\Delta\boldsymbol x_k$, as shown in Figure~\ref{fig: 2d-translation-transition}.

\begin{figure}[H]
    \centering
    \includegraphics[width=0.4\linewidth]{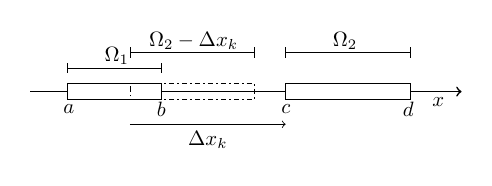}
    \caption{One-dimensional cells $\Omega_1$, $\Omega_2$, and translation $\Omega_2-\Delta x_k$. Under Assumptions \ref{ass: uniform_x} and \ref{ass: ind_step}, the conditional transition probability for a given $\Delta x_k$ corresponds to the proportion of $\Omega_1$ that is also in $\Omega_2-\Delta x_k$, which is hatched.}
    \label{fig: 1d-translation-transition}
\end{figure}

To illustrate, let us calculate transition probabilities between two one-dimensional linear elements $\Omega_1 = [a,b]$ and $\Omega_2 = [c,d]$, as visualised in Figure~\ref{fig: 1d-translation-transition}. We have $V(\Omega_1) = b-a$ and $V(\Omega_2) = d-c$.

One can compute that
\begin{equation}
    V(\Omega_1\cap(\Omega_2-\Delta x_k)) = \begin{cases}
        |\min(b,d-\Delta x_k) - \max(a,c-\Delta x_k)| & \mathrm{if} \quad c-b\leq \Delta x_k \leq d-a \\
        0                                             & \mathrm{otherwise}
    \end{cases}
\end{equation}
and hence
\begin{equation}
    P(x_{k+1} \in \Omega_2 | x_k\in \Omega_1, \Delta x_k) = \begin{cases}
        \frac{|\min(b,d-\Delta x_k) - \max(a,c-\Delta x_k)|}{b-a} & \mathrm{if} \quad c-b\leq \Delta x_k \leq d-a \\
        0                                                         & \mathrm{otherwise.}
    \end{cases}
\end{equation}
Substituting into \eqref{eq:transition-prob} and evaluating the integral yields the sought-after transition probability.

\begin{remark}
    We believe that practical formulas for the computation of transition probabilities between two cells, as the ones given above for the one-dimensional case, should be obtainable for higher-dimensional grid cells as well, provided that the cells are part of a regular grid, where all cells are translations of each other. More specifically, if $\Omega_2 = \Omega_1 + \boldsymbol \alpha$, one could reuse the transformation to local coordinates from Section~\ref{sec: local_coords} and computations in Section~\ref{sec:cond_escp_prob} in this alternative framework to obtain the transition probability between any two cells in a regular grid.
\end{remark}

\newpage
\bibliographystyle{abbrv}
\bibliography{Bib_IngelaereMaes}

\end{document}